\documentclass[onecolumn,showkeys,amsmath,amssymb,amsfonts,a4paper,aps,prd,10pt]{revtex4}
\usepackage[colorlinks, linkcolor={blue},citecolor={red}]{hyperref}
\usepackage{graphicx}
\usepackage{epstopdf}
\newcommand{\nc}{\newcommand}
\nc{\ba}{\begin{eqnarray}} \nc{\ea}{\end{eqnarray}}
\newcommand\be{\begin{equation}}
\newcommand\ee{\end{equation}}
\nc{\D}{\overline{\mbox{D3}}}

\nc{\ga}{\gamma} \nc{\tnu}{\tilde{\nu}} \nc{\tmu}{\tilde{\mu}}

\nc{\x}{{\bf{x}}}

\input{epsf.sty}

\begin{document}
\title{A Conformally Invariant Theory of Gravitation in\\ Metric Measure Space}
\author{Nafiseh Rahmanpour}
\email{n$_$rahmanpour@sbu.ac.ir}
\author{Hossein Shojaie}
\email{h-shojaie@sbu.ac.ir}
\affiliation{Department of Physics, Shahid Beheshti University, G.C., Evin, Tehran 1983969411, Iran}

\begin{abstract}
In this manuscript, a conformally invariant theory of gravitation in the context of metric measure space is studied. The proposed action is invariant under both diffeomorphism and conformal transformations. Using the variational method, a generalization of the Einstein equation is obtained, wherein the conventional tensors are replaced by their conformally invariant counterparts, living in metric measure space. The invariance of the geometrical part of the action under a diffeomorphism leads to a generalized contracted second Bianchi identity. In metric measure space, the covariant derivative is the same as it is in the Riemannian space. Hence, in contrast to the Weyl space, the metricity and integrability are maintained. However, it is worth noting that in metric measure space the divergence of a tensor is not simply the contraction of the covariant derivative operator with the tensor that it acts on. Despite the fact that metric measure space and integrable Weyl space, are constructed based on different assumptions, it is shown that some relations in these spaces, such as the contracted second Bianchi identity, are completely similar.
 
\end{abstract}
\maketitle

\section{Introduction}

In recent years, \textit{metric measure space} (\textit{Riemann measure space}) has had a leading role in some important issues in mathematics~\cite{Perelman, Chang, Case, Case2, Chang2, Chang3, Cheeger, Lott, Sturm 1, Sturm 2}. In this space, in addition to the metric $g$, there is a scalar density function $f$ on the manifold, and the volume element is modified to $dm=\exp{(-f)}dvol(g)$. As the most important application of this space in topology, one can recall the elegant proof of the Poincar\'e conjecture by Perelman~\cite{Perelman}. Chang, Gursky and Yang have studied metric measure space with the tools of conformal geometry and provided a way to construct conformally invariant geometrical objects~\cite{Chang}. Also Case~\cite{Case, Case2}, by using the notion of conformally warped manifolds, has demonstrated how one can unify the two approaches of~\cite{Chang} and~\cite{Lott, Sturm 1,Sturm 2}. 

In the physics literature, on the other hand, Weyl geometry is a well-known framework to construct conformally invariant (Weyl invariant) objects. In Weyl geometry, in addition to the metric, there is a 1-form $\kappa$ on the manifold, which makes the length of a vector change under displacement~\cite{Weyl1}. Non-metricity and non-integrability of length are two important properties of this geometry. 

Based on this geometry, Weyl constructed a conformally invariant action to describe both gravity and electromagnetism~\cite{Weyl2}. In spite of interesting features, Weyl theory suffers from inconsistency with observations due to non-integrability of length. In addition, Weyl had to use squared Ricci scalar as the gravitational part of the action to make it conformally invariant. 

Dirac used Weyl geometry as a framework to construct a theory compatible with his hypothesis about the constants of nature, known as large number hypothesis (LNH)~\cite{Dirac}. LNH predicts a varying gravitational constant G, while general relativity (GR) considers it as a universal constant. To construct a conformally invariant action in Weyl-Dirac theory, a new scalar field $\beta$, in addition to the existing fields in Weyl geometry, was proposed. Dirac related this gauge field to the notions of gravitational (Einstein) and atomic units and used LNH to determine the relationship between these two. He also fixed the integrability of the length in the atomic unit, consistent with observations, while leaving the length non-integrable in the gravitational unit. 

There is another way to resolve the problem of non-integrability in Weyl geometry, by assuming the 1-form $\kappa$ to be the gradient of a scalar field. This is the so-called integrable Weyl geometry. Moreover, by setting $\kappa=\frac{\nabla \beta}{\beta}$, the integrable Weyl-Dirac theory is obtained~\cite{Canuto}.

Conformal transformations appear in several areas of theoretical physics, particularly in gravitational theories, ranging from modified theories of gravitation to AdS/CFT correspondence~\cite{sasaki, Quiros2, Ads1, Witten, Faci, fulton, Turok, Drechsler1, Drechsler2, Manheim1, Manheim2, Manheim3}. The idea of expressing basic laws of physics covariantly under the widest possible group of transformations is one of the primary reasons to regard conformal transformations. Another reason for considering conformal transformations in physical laws is that, it may be interpreted as unit transformations~\cite{Dicke,James}. There are also a lot of controversial discussions about the equivalence of conformally related scalar-tensor theories of gravitation; see for instance~\cite{Flanagan, Quiros2} and the references therein.

The significant role of metric measure space in mathematics, in addition to its relation to conformal transformations which is proposed in~\cite{Chang}, raises hopes for metric measure space to find its way in physics. The aim of this work is to propose a conformally invariant theory of gravitation in the context of metric measure space, as well as to compare it with the well-known conformally invariant theories presented in the context of Weyl geometry.   

It should be noted that, a space with a non canonical volume element (measure) has been also discussed by Abraham et al. as volume manifolds~\cite{Abraham}. This notion was also generalized by Graf, who used volumetric manifold to formulate dilaton gravity~\cite{Graf1} and Ricci flow gravity~\cite{Graf2}.   

This manuscript is arranged as follows: in Sec.~2, the definition of metric measure space is introduced, and some aspects of this space are discussed. The method of Chang et al.~\cite{Chang} to construct conformally invariant geometrical objects is also presented in this section. In the 3rd section, a gravitational action is proposed, and dynamical equations are derived. Identities resulted from the invariance of the action under diffeomorphism and conformal transformation are obtained in Sec.~4. In Sec.~5, after a brief introduction of Weyl space, a comparison between metric measure space and Weyl space is provided. In addition, it is investigated that under what conditions our gravitational action coincides with the action presented in~\cite{Canuto}, as a typical action in the context of integrable Weyl-Dirac geometry. A conclusion and remarks are the aims of the last section. 
\section{Metric Measure Space}

By definition, a metric measure space is a triple ($M^n,g,m$), consisting of a (Pseudo) Riemannian manifold ($M^n, g$) and a smooth measure $m$. The density function $f$ associated to $m$ is defined by 
\begin{equation}\label{0}
dm=\exp{(-f)} dvol(g). 
\end{equation}
In the mathematical point of view, the name ``density function'' shows that $f$ can be interpreted as a weight or a probability function on the manifold. However, in a physical point of view, one may consider it as a function used to scale units at each point of the manifold.

In the following subsections, it is shown that how a geometrical object in metric measure space may be defined to be conformally invariant. The concept of weighted divergence in this space is also discussed. 
\subsection{Conformally invariant geometrical objects in metric measure space}

In the approach of Chang et al.~\cite{Chang}, the dependence of conformally invariant objects on measure $m$ is mediated by the density function $f$, defined in (\ref{0}). In this setting, every Riemannian invariant $I(g)$ of $(M^n,g)$, gives rise to a conformal density of weight $s$ of metric measure space ($M^n,g,m$), denoted by $\mathcal{I}_s(g,f)$. The conformal density $\mathcal{I}_s(g,f)$ has two properties: $\mathcal{I}_s(g,0)=I{(g)}$ and $\mathcal{I}_s(\hat{g},\hat{f})=\exp{(sw)}\mathcal{I}_s(g,f)$ where $\hat{g}=\exp{(2w)}g$, and $\hat{f}$ denotes the density function associated to $\hat{g}$. 

Before proceeding to construct $\mathcal{I}_s(g,f)$ from $I(g)$, it should be determined how the density function $\hat{f}$ is related to density function $f$. Under a conformal transformation $\hat{g}=\exp{(2w)}g$, one has $dvol(\hat{g})=\exp{(nw)} dvol{(g)}$ for an n-dimensional manifold. Hence, if one assumes $d\hat{m}=\exp{(s^\prime w)}dm$, straightforward calculations lead to
 \begin{equation}\label{1}
 \hat{f}=f+(n-s^\prime)w. 
 \end{equation}
It is easy to check that the relation $\mathcal{I}_s(g,f)= \exp{(\frac{fs}{n-s^\prime})} I(\exp{(\frac{-2f}{n-s^\prime})}g)$, satisfies the desired properties of a conformal density of weight $s$, when the joint transformations $\hat{g}=\exp{(2w)}g$ and $\hat{f}=f+(n-s^\prime)w$ are applied. 

As it is clear from (\ref{1}), there is an arbitrariness in the way that the density function $f$ is transformed under a conformal transformation, depending on the weight of the measure $dm$. For instance, Chang et al.~\cite{Chang} assume the measure $dm$ is conformally invariant, i.e. $d\hat{m}=dm$, while Case~\cite{Case} assumes $d\hat{m}=\exp{(2w)}dm$. The different choice of the weight of measure $dm$, leads to different coefficients of derivatives of $f$ in $\mathcal{I}(g,f)$ and seems to be a matter of convention.  

In this way, the generalized Ricci tensor and scalar with the corresponding weights $s_1$ and $s_2$ on an $n$-dimensional manifold are
are
\begin{eqnarray}\label{3}
\mathcal{R}_{\mu\nu}& =& \exp{(\frac{fs_1}{n-s^\prime})}[ R_{\mu\nu} + \frac{n-2}{n-s^\prime}\nabla_\mu \nabla_\nu f +\frac{n-2}{(n-s^\prime)(n-s^\prime)}\nabla_\mu f \nabla_\nu f \nonumber \\
&+& \frac{1}{n-s^\prime}\nabla^\alpha \nabla_\alpha f g_{\mu\nu} -\frac{n-2}{(n-s^\prime)(n-s^\prime)}\nabla^\alpha f \nabla_\alpha f g_{\mu\nu}]
\end{eqnarray}
and
\begin{eqnarray}\label{2}
\mathcal{R} = \exp{(\frac{s_2 f+2f}{n-s^\prime})}[R + 2(\frac{n-1}{n-s^\prime})\nabla^\mu \nabla_\mu f -\frac{(n-1)(n-2)}{(n-s^\prime)(n-s^\prime)}\nabla^\mu f \nabla_\mu f],
\end{eqnarray}
where $R_{\mu\nu}$ and $R$ are the traditional Ricci tensor and scalar in (Pseudo) Riemannian space, respectively. These traditional tensors are retrieved from their conformally invariant counterparts, by setting $f=0$. Since $\mathcal{R}=g^{\mu\nu}\mathcal{R}_{\mu\nu}$, the weight of $\mathcal{R}$ is the sum of the weights of the Ricci tensor and the inverse of the metric, i.e. $s_2=s_1-2$. It is evident that the choice $s_1=0$ makes the coefficients of both the traditional Ricci tensor and scalar, in equations (\ref{3}) and (\ref{2}), be unity. As it will be shown in Sec.~5, this choice is also a necessary condition for (\ref{3}) and (\ref{2}) to be in accordance with their counterparts in integrable Weyl geometry. In addition, in Sec.~3 we will argue that, by demanding the proposed action (\ref{4}) to be conformally invariant, one gets $s^\prime=-s_2$ for the weight of measure $dm$. Therefore, in this manuscript, we prefer to set $s_1=0$ (or equivalently $s_2=-2$) and $s^\prime=2$, as well.
 
\subsection{Divergence}
One of the natural notions in metric measure space is the concept of the weighted divergence~\cite{Chang2,Case}. The definition of divergence in mathematics is mainly represented in two ways, the divergence as the Lie derivative of a density and the divergence as the adjoint of the covariant derivative. 

On an n-dimensional orientable manifold, a density can be identified with an n-form. Hence, in the first approach, the divergence of a vector field $X=X^\mu \partial_\mu$, with respect to the n-form $\varpi= \omega dx^1\wedge ....\wedge dx^n$, is defined as 
\begin{equation}\label{1a}
(div_\varpi X) \varpi= L_X \varpi.  
\end{equation}
To have more familiar form of divergence, it is useful to use Cartan's relation, namely   
\begin{equation}\label{2a}
L_X \varpi=d i_{X}\varpi+i _X d\varpi,
\end{equation}
where $i_{X}$ and $d$ are the interior product and exterior derivative, respectively. For n-form $\varpi$ the equation (\ref{2a}) reduces to  
\begin{equation}\label{3a}
L_X \varpi=d i_{X}\varpi,
\end{equation}
since on an n-dimensional manifold one has $d\varpi=0$. The right-hand side of (\ref{3a}), in terms of local coordinate, is
\begin{equation}\label{4a}
d i_{X}\varpi= \partial_\mu(\omega X^\mu) dx^1\wedge .....\wedge dx^n.
\end{equation}
Therefore, regarding (\ref{1a}), (\ref{3a}) and (\ref{4a}) together, one gets
\begin{equation}\label{5a}
div_\varpi X = \frac{1}{\omega} \partial_\mu(\omega X^\mu).
\end{equation}
If $X$ is a vector field on a manifold endowed with metric $g$, it is natural to consider $\varpi$ as the standard volume form, namely $\eta\equiv\varpi=\sqrt{-g} dx^1\wedge ....\wedge dx^n$. With this choice, and using $\Gamma^\mu_{\mu\nu} =\frac{1}{\sqrt{-g}}\partial_\nu (\sqrt{-g})$, it can be shown that (\ref{5a}) becomes
\begin{equation}\label{6a}
div_\eta  X  = \nabla_\mu X^\mu,
\end{equation}
which is the familiar relation for divergence in Riemannian geometry and consequently in the context of GR. 

In metric measure space, however, the volume form is defined as $\zeta\equiv\varpi=\exp{(-f)}\sqrt{-g} dx^1\wedge ....\wedge dx^n$ and hence, the divergence of a vector field $X$ is
\begin{equation}\label{7a}
 div X \equiv div_\zeta X =\nabla_\mu X^\mu-\nabla_\mu f X^\mu. 
\end{equation}

In the second approach, divergence is defined through an integral relation, namely as an adjoint of covariant derivative~\cite{Chang2}. For instance, divergence of a symmetric (0,2) tensor $Y$ is defined as
\begin{equation}\label{7s}
\int \langle Y, \nabla \vartheta \rangle dm = -\int \langle div Y, \vartheta \rangle dm,
\end{equation}
where $\vartheta$ is a 1-form. Straightforward calculations show that this relation in metric measure space, where $dm$ is supposed as (\ref{0}), leads to   
\begin{equation}\label{7n}
div Y= \nabla_{\mu} Y^{\mu\nu}-\nabla_\mu f Y^{\mu\nu}.
\end{equation}

Therefore, as it is clear from these two approaches, in metric measure space, divergence is not equal to the contracted covariant derivative anymore. In addition, divergence of a tensor field of an arbitrary rank is derived similarly. For instance, the divergence of a contravariant tensor field $T$ in metric measure space is 
\begin{equation}
div T= \nabla_\mu T^{\mu\nu....} -\nabla_\mu f T^{\mu\nu....}. 
\end{equation} 
\section {Lagrangian}
In this section, we propose a gravitational action in the context of metric measure space. A simple conformally invariant extension of the Einstein-Hilbert action in this space can be
\begin{equation} \label{4}
S= \int \{\frac{1}{2\kappa} (\mathcal{R}- 2\exp{(-f)}\Lambda)+\mathcal{L}_{matter}(g,f, \psi)\} dm,
\end{equation}
where $\psi$ is the matter field, and $\kappa$ as well as $\Lambda$, are pure constants. Since we have assumed the weight of $\mathcal{R}$ to be $s_2=-2$, then for the action to be conformally invariant, the weight of $dm$ should be $s^\prime=2$. Hence, with the same reasoning, $\mathcal{L}_{matter}$ will be a conformal density of weigh $-2$. The constants $\Lambda$ and $\kappa$ are assumed to be weightless. 

Variations with respect to the metric $g$ and the density function $f$ yield
\begin{equation}\label{5}
\frac{1}{2\kappa}(\mathcal{R}_{\mu\nu}-\frac{1}{2} \mathcal{R} g_{\mu\nu})+\frac{1}{2\kappa}(\exp{(-f)}\Lambda g_{\mu\nu})+\frac{1}{\sqrt{-g}}\frac{\delta(\sqrt{-g}\mathcal{L}_{matter})}{\delta g_{\mu\nu}}=0
\end{equation}
and \begin{equation}\label{6}
\frac{-1}{2\kappa}(\mathcal{R})+\frac{1}{2\kappa}(4\exp{(-f)}\Lambda)+\frac{1}{\exp{(-f)}}\frac{\delta(\exp{(-f)}\mathcal{L}_{matter})}{\delta f}=0
\end{equation}
respectively, with both equations to be conformally invariant. Defining the stress-energy tensor as
\begin{equation}\label{6-1}
\mathcal{T}_{\mu\nu}=\frac{-2}{\sqrt{-g}}\frac{\delta(\sqrt{-g}\mathcal{L}_{matter})}{\delta g_{\mu\nu}} ,
\end{equation}
equation (\ref{5}) can be rewritten as
\begin{equation}\label{10}
\mathcal{R}_{\mu\nu}-\frac{1}{2} \mathcal{R} g_{\mu\nu}+\exp{(-f)}\Lambda g_{\mu\nu} =\kappa \mathcal{T}_{\mu\nu}.
\end{equation}
 
On the other hand, and as it will be shown in the next section, the identity derived from the conformal invariance of the action (\ref{4}) yields
\begin{equation}\label{9}
\mathcal{T}=\frac{-2}{\exp{(-f)}}\frac{\delta(\exp{(-f)}\mathcal{L}_{matter})}{\delta f},
\end{equation}
where $\mathcal{T}$ is the trace of the stress-energy tensor $\mathcal{T}_{\mu\nu}$. Comparing the trace of equation (\ref{10}) with using equation (\ref{9}), shows that equations (\ref{5}) and (\ref{6}) are not independent, and the only independent dynamical equation is (\ref{10}). This equation is similar to the Einstein equation. However, each quantity including the Einstein and the stress-energy tensors, in addition to the cosmological constant term, have been replaced by their conformally invariant counterparts living in metric measure space.

By setting $f=0$ in (\ref{10}), one gets
\begin{equation}\label{20}
R_{\mu\nu}-\frac{1}{2} R g_{\mu\nu}=-\Lambda g_{\mu\nu}+\kappa T_{\mu\nu}.
\end{equation}
This special conformal frame can be dubbed as the Einstein frame. To retrieve exactly the classical Einstein equation in this frame and respecting the correspondence principle, one has to set $\kappa=8\pi$. 

It is worth noting that according to Chang et al.~\cite{Chang}, a fixed $m$ under diffeomorphism leads to an interesting result about the Yamabe problem. Nevertheless, with this assumption, the gravitational equation does not reduce to the Einstein equation in the case of $f=0$. Hence, according to the importance of the correspondence principle in physics, we do not follow~\cite{Chang} in fixing $m$ under diffeomorphism. Moreover, the geometrical part of the action (\ref{4}) is a special case of energy functional $\mathcal{W}$, defined in~\cite{Case} on a 4-dimensional conformal warped manifold and, therefore, the variational relations derived here, are the special forms of those of~\cite{Case}.

\section{Identities}
The action (\ref{4}) is invariant under both diffeomorphism and conformal transformations, and the invariance under each of these transformations leads to an identity. In this section, two types of identities are derived via the invariance of the action (\ref{4}) under these transformations. It is also shown that how conservation of the stress-energy tensor is considered in metric measure space. 

\subsection{Identities resulting from diffeomorphism invariance}
Let's consider the geometrical action
\begin{equation}\label{18i}
S_{g}=\frac{1}{2\kappa} \int(\mathcal{R}- 2\exp{(-f)}\Lambda)dm, 
\end{equation}
with corresponding variation 
\begin{equation}\label{18}
\delta S_{g}=\frac{1}{2\kappa}\int \{(\mathcal{R}_{\mu\nu}-\frac{1}{2} \mathcal{R} g_{\mu\nu}+\exp{(-f)}\Lambda g_{\mu\nu}) \delta g^{\mu\nu}+(-\mathcal{R}+4\exp{(-f)}\Lambda)\delta f \} dm.
\end{equation}
For a diffeomorphism to be a symmetry of the action (\ref{18i}), one should have
\begin{equation}
\delta S_{g}=0.
\end{equation}
In addition, $\delta g^{\mu\nu}= L_X g^{\mu\nu}$ and $\delta f=Xf$ are hold for a diffeomorphism generated by a vector field $X$. Substituting $L_X g^{\mu\nu}=-(\nabla^\mu X^\nu+\nabla^\nu X^\mu)$ and $X f=X^\nu \nabla_\nu f$, and paying attention that $R_{\mu\nu}$ is symmetric, one obtains
\begin{equation}\label{17g}
\int \{-2(\mathcal{R}_{\mu\nu}-\frac{1}{2} \mathcal{R} g_{\mu\nu}+\exp{(-f)}\Lambda g_{\mu\nu}) \nabla^\mu X^\nu+(- \mathcal{R}+4\exp{(-f)}\Lambda)X^\nu\nabla_\nu f\}dm=0.
\end{equation}
Regarding the definition of divergence (\ref{7s}), the relation (\ref{17g}) can be written as 
\begin{equation}
\int \{2 div (\mathcal{R}_{\mu\nu}-\frac{1}{2} \mathcal{R} g_{\mu\nu}+\exp{(-f)}\Lambda g_{\mu\nu}) X^\nu+(- \nabla_\nu f\mathcal{R}+4\nabla_\nu f \exp{(-f)}\Lambda)X^\nu\} dm=0.
\end{equation}
Ultimately, by applying divergence, according to (\ref{7n}), one gains the relation
\begin{equation}\label{16}
\nabla_{\mu} \mathcal{R}^{\mu\nu}-\nabla_{\mu}f \mathcal{R}^{\mu\nu} -\frac{1}{2}\nabla^{\nu} \mathcal{R} =0,
\end{equation}
as the generalized contracted second Bianchi identity in metric measure space. This result is in accordance with the contracted second Bianchi identity provided in~\cite{Chang2}.

Note that, when divergence is applied to a scalar, it acts as the covariant derivative. Indeed by (\ref{5a}), it is easy to check that the relation
\begin{equation}
div (\mathcal{R}g)= \mathcal{R}div (g)+g\nabla \mathcal{R},
\end{equation}
is satisfied for any tensor multiplied by a scalar, particularly here $g$ and $\mathcal{R}$.

Similarly, from the matter action 
\begin{equation}\label{16hh}
 S_{m}= \int \mathcal{L}_{matter}(g,f, \psi) dm,
\end{equation}
one has 
\begin{equation}\label{16sd}
\nabla_{\mu}\mathcal{T}^{\mu\nu}-\nabla_\mu f\mathcal{T}^{\mu\nu}+\frac{1}{2}\nabla^\nu f \mathcal{T}=0,
\end{equation}
where (\ref{16sd}) can be considered as a generalization of the traditional relation $\nabla_\mu T^{\mu\nu}=0$.  

\subsection{Identity resulting from conformal invariance}
In the case of conformal transformations, the relations $\delta g=2wg$ and $\delta f=2w$ are hold. Substituting these in
\begin{equation}
\delta S=0,
\end{equation}
one gets
\begin{equation}\label{19}
\mathcal{T}=\frac{-2}{\exp{(-f)}}\frac{\delta(\exp{(-f)}\mathcal{L}_{matter})}{\delta f}.
\end{equation}

As mentioned in Sec.~3, this identity leads equation (\ref{10}) to be the only dynamical equation. Indeed, equation (\ref{10}) together with the Bianchi identity (\ref{16}), provide six independent equations, where there are eleven unknown functions, i.e. ten for the metric components, and one for the scalar density $f$. While four components of the metric can be fixed by coordinate conditions as in GR, there remains an undetermined function $f$. 

The origin of this under-determinacy is the invariance of the action under a conformal transformation and should come as no surprise.
There are detailed discussions about the physical interpretations of this type of freedom and methods of fixing it, from both theoretical and phenomenological viewpoints, such as topological considerations, Mach principle and LNH~\cite{Canuto, Canuto3, Canuto2}. For instance, in the work of Canuto et al.~\cite{Canuto}, where a conformally invariant theory of gravitation in the context of integrable Weyl-Dirac geometry is provided, an identity similar to (\ref{19}) gives rise to under-determinacy of the Dirac gauge field $\beta$. There, they suggested to apply LNH as a way to fix the so-called gauge freedom. However, proposing a particular way for determining this freedom is not in the scope of this manuscript, and we do not offer a dynamical equation for $f$. 
 
\section{connection with Integrable Weyl geometry}

Weyl geometry is considered as the natural candidate for extending the Riemannian geometry to construct conformally invariant geometrical objects. This geometry, although not very often, is used in the physics literature and cosmology~\cite{Rosen, Erhard1, Erhard, Miritzis, Shojai, R-Carroll, S1, S2, S3}. It seems that there are similar and different aspects between metric measure space and integrable Weyl space. To compare these aspects, it is useful to introduce Weyl geometry in brief. 

\subsection{A glance at Weyl space}
In this subsection, the essential features of Weyl geometry will be reviewed. Weyl space is a manifold endowed with a metric $g$ and a 1-form $\kappa$. The covariant derivative $\overline{\nabla}$, is defined through the non-metricity condition  
\begin{equation}\label{21a}
\overline{\nabla}_\lambda g_{\mu\nu}= -2\alpha \kappa_\lambda g _{\mu\nu},
\end{equation}
in which $\alpha$ is an arbitrary constant. The covariant derivative $\overline{\nabla}$, has connection components
\begin{equation}\label{21b}
\overline{\Gamma}^\rho_{\mu\nu}= \Gamma^{\rho}_{\mu\nu}+C^{\rho}_{\mu\nu},
\end{equation}
where $\Gamma^{\rho}_{\mu\nu}$ is the Levi-Civita connection and
\begin{equation}
C^{\rho}_{\mu\nu}=\alpha (\delta^\rho_\nu \kappa_{\mu} +\delta^\rho_\mu \kappa_\nu -g_{\mu\nu} \kappa^{\rho}).
\end{equation}
The connection $\overline{\Gamma}^\rho_{\mu\nu}$ is invariant under the joint transformations $\hat{g}_{\mu\nu}=\exp{(2w)}g_{\mu\nu}$ and $ \hat{\kappa}_\mu=\kappa_\mu-\frac{1}{\alpha}\nabla_\mu w$. 

The condition (\ref{21a}) leads to change of length under parallel displacement and hence the non-integrability of length. In Weyl space, conformally invariant geometrical objects are constructed by the generalized covariant derivative $\overline{\nabla}$. Consequently, the corresponding Riemann tensor $\overline{R}^\rho_{\mu\nu\lambda}$ is no longer antisymmetric in it's first two indices and the Ricci tensor $\overline{R}_{\mu\nu}$ is not symmetric, as well. However, if one considers 1-form $\kappa$ as a gradient of some scalar field $\varphi$, the integrability of length is retrieved, and the Ricci tensor becomes symmetric. This modified model is called the integrable Weyl geometry. The conformally invariant Ricci tensor and Ricci scalar in integrable Weyl geometry are
\begin{equation}\label{22f}
\overline{R}_{\mu\nu}= R_{\mu\nu} - (n-2)\alpha \nabla_\mu \kappa_\nu +(n-2)\alpha^2 \kappa_\mu  \kappa_\nu  - \alpha \nabla^\sigma \kappa_\sigma  g_{\mu\nu} -(n-2)\alpha^2 \kappa^\sigma \kappa_\sigma  g_{\mu\nu}
\end{equation}
and 
\begin{equation}\label{23s}
\overline{R}= R-2(n-1)\alpha \nabla_\mu \kappa^\mu-(n-2)(n-1) \alpha^2 \kappa _\mu \kappa^\mu.
\end{equation}
Note that the Ricci tensor (\ref{22f}) and the Ricci scalar (\ref{23s}) have the conformal weights 0 and -2, respectively. The contracted second Bianchi identity in integrable Weyl geometry is different from that of Riemannian space and is given by 
\begin{equation}\label{32a}
\overline{\nabla}_\mu \overline{G}^{\mu}_{\nu}-2\alpha \kappa_\mu \overline{G}^{\mu}_{\nu}=0,
\end{equation}
where $\overline{G}^{\mu\nu}$ is the conformally invariant Einstein tensor. If one writes the identity (\ref{32a}) in terms of  the covariant derivative $\nabla$, it reduces to
\begin{equation}\label{32r}
\nabla_\mu \overline{R}^{\mu\nu} -\frac{1}{2}\nabla^\nu \overline{R} +2\alpha \kappa_\mu \overline{R}^{\mu\nu}=0.
\end{equation}
The aim of the next subsection is to introduce three of the well-known works in the context of Weyl geometry.  

\subsection{Actions provided by Weyl, Dirac and Canuto}
In the literature, there have been variety of actions within the framework of Weyl geometry. Weyl himself, in 1918, proposed a theory to describe electromagnetism as well as gravitation. Weyl's action is 
\begin{equation}
I=\int (\frac{1}{4} F_{\mu\nu} F^{\mu\nu}+\overline{R}^{2}) dvol(g),
\end{equation}
where $F_{\mu\nu}=\partial_\mu\kappa_\nu-\partial_\nu\kappa_\mu$ is the electromagnetic field tensor. Since the weight of the Ricci scalar $\overline{R}$ in Weyl geometry is -2, he used the square of the Ricci scalar to construct a conformally invariant action. In addition to the inherent problem of non-integrability in Weyl geometry, the derived equations, due to the term $\overline{R}^2$, were complicated. 

Afterwards, in 1973, Dirac proposed an action, much simpler than that of Weyl, but with an extra scalar field $\beta$~\cite{Dirac}. Indeed, the Dirac field $\beta$ with the conformal weight -1, helps one construct a conformally invariant action in which the Ricci scalar appears at first order. Moreover, he used the notion of the atomic and the gravitational clocks to justify the non-integrability of Weyl space. Dirac's action is 
\begin{equation}
I=\int (\frac{1}{4} F_{\mu\nu}F^{\mu\nu} - \beta^2 \overline{R} +c_1\beta^{\ast\mu} \beta_{\ast\mu}+c_2\beta^4) dvol(g), 
\end{equation}
where $c_1$ and $c_2$ are two arbitrary constants and $\beta^{\mu\ast}$ is defined as the co-covariant derivative of $\beta$. In Dirac's theory, there are three independent fields $g$, $\kappa$ and $\beta$. The variation of the action with respect to these fields gives rise to three equations. Dirac set $c_1=6$ and dropped the term $c_2 \beta^4$, since it would be important only in cosmology~\cite{Dirac}. The field equation corresponding to the variation of $\beta$ is the trace of the derived equation from the variation of $g$ and hence, they are not independent. In addition, the field equation resulted from the variation of $\kappa$ gives rise to the Maxwell equations. 

Canuto et al., in 1977, followed Dirac's idea with a minor modification, and considered the theory in the context of integrable Weyl geometry by assuming $\kappa=-\nabla(\ln \beta)$ ~\cite{Canuto}. As a result, there remain just  two fields in the theory, namely $g$ and $\beta$, and the term $F_{\mu\nu} F^{\mu\nu}$ disappears from the action. The choice $\kappa=-\nabla(\ln \beta)$ implies $ \beta^{\ast\mu}=0$, as well. Beside setting $\overline{\Lambda}=\frac{1}{2} c_2 \beta^2$, they added the Lagrangian of matter $\overline{L}_{matter}(g,\psi,\beta)$, and proposed
\begin{equation}\label{30b}
I=\int (- \beta^2 \overline{R} +2\overline{\Lambda} \beta^2 +16\pi \overline{L}_{matter}) dvol(g),
\end{equation}
as a gravitational action in integrable Weyl geometry. 

In the next subsection, we will show how some relations in metric measure space and integrable Weyl space can be the same. Also, by comparing the actions (\ref{4}) and $(\ref{30b})$, we find a relation between the density function $f$ and the Dirac field $\beta$. 


\subsection{Comparing with metric measure space}
As mentioned in Sec.~2, in metric measure space, one has the joint transformations $\hat{g}=\exp{(2w)}g$ and $\hat{f}=f+(n-s^\prime)w$, and the conformally invariant Ricci tensor and Ricci scalar of weights $s_1$ and $s_2$ are given by (\ref{3}) and (\ref{2}), respectively. To clarify the similarities between integrable Weyl space and metric measure space, let's consider $\kappa$ as 
\begin{equation}\label{23}
\kappa_\mu=\gamma\nabla_\mu f,
\end{equation} 
where $f$ is the density function, defined in metric measure space, and $\gamma$ is a constant. Comparing the two transformations $\hat{f}=f+(n-s^\prime)w$ and $ \hat{\kappa}_\mu=\kappa_\mu-\frac{1}{\alpha}\nabla_\mu w$, one finds that these relations will be similar, provided that
\begin{equation}\label{23k}
\frac{1}{n-s^\prime}=-\alpha \gamma. 
\end{equation}
Moreover, the generalized Ricci tensor and scalar in metric measure space, relations (\ref{3}) and (\ref{2}), will be the same as those in integrable Weyl space, relations (\ref{22f}) and (\ref{23s}), provided that $s_1=0$. In other words, assuming (\ref{23k}) and $s_1=0$, (or equivalently $s_2$=-2), leads to
\begin{equation}\nonumber
\mathcal{R}_{\mu\nu}=\overline{R}_{\mu\nu},
\end{equation}
and
\begin{equation}\nonumber
\mathcal{R}=\overline{R}.
\end{equation}

As explained in Sec.~3, the choice $s_1=0$ as well as the form of the proposed action (\ref{4}), set $s^\prime=2$ as the weight of measure in 4-dimensional metric measure space. Therefore, to compare the derived results of our theory (action) in metric measure space, with their counterparts in integrable Weyl geometry, one should consider $\gamma=-\frac{1}{2\alpha}$. One of these result is the contracted second Bianchi identity (\ref{16}). By the mentioned assumptions, it is easy to show that the identity (\ref{16}) coincides with its counterpart in integrable Weyl geometry, relation (\ref{32r}).

In the last subsection, we introduced three important actions in Weyl space. Here, we compare the action (\ref{4}) in metric measure space, with that of Canuto et al, the action (\ref{30b}), as a typical gravitational action in the context of integrable Weyl geometry. Before proceeding further, note that Canuto et al.~\cite{Canuto} (and also Dirac~\cite{Dirac}), set $\alpha=-1$. Hence in comparing (\ref{4}) and $(\ref{30b})$, one should consider $\gamma=\frac{1}{2}$, and consequently $\kappa=\frac{1}{2}\nabla f$. In addition, they assumed that $\kappa=-\nabla(\ln \beta)$, which leads to 
\begin{equation}
\beta^2=\exp{(-f)}. 
\end{equation}
Therefore, the action (\ref{30b}) can be rewritten as 
\begin{equation}\label{33f}
I=\int \{\frac{1}{16\pi}(\overline{R} -2\overline{\Lambda}) - \exp{(f)} \overline{L}_{matter}\} dm,
\end{equation}
where $dm=\exp{(-f)} dvol(g)$. Comparing the actions (\ref{4}) and $(\ref{30b})$, term by term, one finds
\begin{equation}\label{35r}
\overline{\Lambda}=\exp{(-f)}\Lambda
\end{equation}
and
\begin{equation}\label{35t}
\overline{L}_{matter}=-\exp{(-f)}\mathcal{L}_{matter}. 
\end{equation} 
Relation (\ref{35r}) indicates that two constants of the theories, namely $\Lambda$ and $c_2$, only differ by a coefficient of $\frac{1}{2}$, recalling $\overline{\Lambda}=\frac{1}{2} c_2 \beta^2$. Relation (\ref{35t}) demonstrates the difference in the conformal weight of the Lagrangian of matter in these theories. Besides, the minus sign in the RHS of relation (\ref{35t}) is just due to the different sign conventions in the theories. It should be noted that we could write the Lagrangian of matter in the action (\ref{4}), as
 $$S_m=\int \exp{(f)}\mathcal{L}_{matter}^{\prime} dm,$$ 
where $\mathcal{L}_{matter}^{\prime}$ would be a conformal density of weight $-4$, similar to $\overline{L}_{matter}$. 

Putting all these results together, the equivalence of the dynamical equations of these two theories is evident. In addition, as it has been noted, the conformal invariance of the actions in both theories, implies that the equation derived from the variation of the scalar field is the trace of the equation derived from the variation of metric. 

We emphasize that although different assumptions in metric measure space and integrable Weyl space are used, some of the resulting relations are completely similar.
   
\section{conclusion and Remarks}

In this manuscript, a conformally invariant theory of gravitation in the context of metric measure space is studied. Within this new framework, one can construct conformally invariant gravitational action by the two fields, the metric $g$, and the density function $f$. In the proposed action, the Ricci scalar $R$ and the canonical volume element $dvol(g)$ are respectively replaced by the conformally invariant Ricci scalar $\mathcal{R}$ and the volume element $dm$, defined in metric measure space. The variations of the action with respect to the metric and the density function give two equations. However, the invariance of the action under a conformal transformation shows that these equations are not independent. Eventually, one gets a dynamical equation for gravitation, in which by setting $f=0$, the traditional Einstein equation in Riemann space is retrieved.  

The contracted second Bianchi identity and a generalization of conservation of the stress-momentum tensor are obtained through the diffeomorphism invariance of the action. It is shown that by setting $\kappa=-\frac{1}{2\alpha}\nabla f$, the conformally invariant Ricci tensor and scalar, in addition to the contracted second Bianchi identity in the integrable Weyl geometry and metric measure space, are the same. 

It is worth noting that the well-known passage from the Einstein frame to the Jordan frame in the Brans-Dicke theory of gravitation, can be understood as the passage from Riemannian space to integrable Weyl space~\cite{Quiros2}. That is because, integrable Weyl space is not a genuine Weyl space and its scalar field $\varphi$ can be gauged away by a suitable conformal transformation~\cite{Schouten}. Accordingly, it seems that metric measure space, similar to integrable Weyl space, can be regarded as a framework to discuss about the conformal mappings relating the scalar-tensor theories of gravitation.   

Table (\ref{112}) summarizes some similar and different aspects of metric measure space and some other geometries.

     \begin{table}[h]
\caption{metric measure space compared with the three more common geometries.} \label{112}
    \tiny
     \begin{tabular}{|c|c|c|c|c|}
           \hline
     & Riemannian Geometry & Weyl Geometry & Integrable Weyl Geometry & Metric Measure Space  \\ \hline
       \parbox[t]{2cm}{Metricity\\[1mm]} & Yes & No& No& Yes  \\ \hline
        \parbox[t]{2cm}{Integrability\\[1mm]} & Yes & No& Yes  &Yes  \\ \hline  
     \parbox[t]{3cm}{Conformally Invariant Objects\\[1mm]} & No &  Yes & Yes & Yes \\ \hline
     \parbox[t]{3.5cm}{Divergence as Contraction of the Covariant Derivative\\[1mm]} &Yes & Yes&Yes &No\\ \hline
     
    \end{tabular}
\end{table}

As mentioned in Introduction, conformal symmetry reflects the freedom to locally choose a system of units and conformal transformations is  interpreted as unit transformations. It is believed that there are two preferred system of units, namely the gravitational and the atomic units, which may not necessarily coincide. Any deviation of one unit from the other makes the Einstein equation inappropriate for describing phenomena. This is because the Einstein equation is written in the gravitational unit, while observations are usually analyzed in the atomic unit~\cite{Canuto3, Canuto2}. In theories formulated in the (integrable) Weyl-Dirac framework, the Dirac gauge field $\beta$ determines the relation between these units~\cite{Dirac, Canuto, Canuto2}. On the other hand, in a theory formulated in metric measure space, it seems that one can use $f$ (or a function of it, for instance, $e^{-f}$) to denote the relation of these units on each point of the manifold. Therefore, determining the density function $f$ in this space is similar to determining the gauge field $\beta$ in (integrable) Weyl-Dirac geometry.

The geometry of a manifold in Riemannian space changes under a conformal transformation (unit transformation) and hence, it is not a proper framework to define running units on a manifold~\cite{Dicke}. However, metric measure space, due to the presence of the density function $f$, provides a natural framework to have running units. Indeed, the Ricci tensor $\mathcal{R}_{\mu\nu}$ and the Ricci scalar $\mathcal{R}$, defined in metric measure space, are invariant under conformal transformations and, therefore, the geometry constructed by these tensors admits running units on the manifold.   

The variation of the constants of nature may be considered equivalent to the changes of the standard of unit~\cite{Barrow, Elis, Joao}. Hence, it seems that the varying-constant theories can also be formulated in the context of metric measure space. As a last remark, there are theorems that relate the conformal transformation (Weyl rescaling) to the conformal coordinate transformations on which conformal field theory is based. The authors believe that this framework may be regarded as a new way to unify gravity and other fields in physics.

\section*{Acknowledgment}
We would like to thank Eaman Eftekhary for fruitful and constructive discussions. We also thank Shahram Jalalzadeh and Wolfgang Graf for their useful comments. 

\newpage
\appendix
\section{\\Two simple examples of matter field in metric measure space} \label{AppendixA}

In the context of metric measure space, we have considered the Lagrangian of matter $\mathcal{L}_{matter}$ generally as a function of matter field $\psi$, metric $g$ and the density function $f$. It is constructive to investigate two simple cases of this Lagrangian, namely the vacuum energy $\mathcal{L}_{vac}$ and the electromagnetism $\mathcal{L}_{em}$. 

For the first example, let's consider 
\begin{equation}
\mathcal{L}_{vac}= -\exp{(\frac{-s^\prime f}{n-s^\prime})} \rho_{vac}
\end{equation}
as the Lagrangian of the vacuum in an n-dimensional metric measure space. This Lagrangian is constructed in such a way that the matter action
\begin{equation}
S_{m}=\int \mathcal{L}_{matter} dm
\end{equation}
be conformally invariant. The corresponding stress-energy tensor will be
\begin{equation}\label{bb}
\mathcal{T}_{\mu\nu}= -\exp{(\frac{-s^\prime f}{n-s^\prime})} \rho_{vac} g_{\mu\nu}. 
\end{equation}
Since the cosmological constant $\Lambda$ has the same effect as an intrinsic energy density of the vacuum $\rho_{vac}$, one may have $\rho_{vac} = \frac{ \Lambda}{\kappa}$. Hence, the relation (\ref{bb}) for $s^\prime=2$ and $n=4$ will be the same as the cosmological term in equation (\ref{10}), as one expects. Here, we have considered $\Lambda$ as a constant, however, one may regard the conformally invariant  quintessence models by adding a kinetic term to the Lagrangian.

To investigate the second example, recall that in Riemannian space the action for electromagnetism is given by
\begin{equation}\label{cc}
S_{m}=\int (\frac{1}{4} F^{\mu\nu} F_{\mu\nu}) dvol(g).
\end{equation}
The electromagnetic tensor $F_{\mu\nu}= \partial_\mu A_\nu- \partial_\nu A_\mu$ is defined independently of the metric $g$ and does not change under a conformal transformation $\hat{g}_{\mu\nu}=\exp{(2w)}g_{\mu\nu}$. However, under such a transformation $F^{\mu\nu}=g^{\mu\alpha}g^{\nu\beta} F_{\alpha\beta}$ is transformed as $\hat{F}^{\mu\nu}=\exp{(-4w)} F^{\mu\nu}$. Since $dvol(\hat{g})=\exp{(nw)} dvol(g)$, the action (\ref{cc}) is conformally invariant in four dimensions. 

In metric measure space, on the other hand, $dm$ and $F^{\mu\nu} F_{\mu\nu}$ have the conformal weights $s^\prime$ and $-4$, respectively. To construct a conformally invariant action on a 4-dimensional metric measure space, one should multiply the $F^{\mu\nu}F_{\mu\nu}$ by $\exp{(f)}$ (recalling that $\exp{(f)}$ has the conformal weight $n-s^\prime$ in n dimensions), hence
 \begin{equation}
 \mathcal{L}_{em}=\frac{1}{4}\exp{(f)} F^{\mu\nu}F_{\mu\nu}.
\end{equation}
As a result, the term $\exp{(f)}$ cancels out the $\exp{(-f)}$ in the measure $dm$, and consequently the conformally invariant action for electromagnetism in the context of metric measure space remains the same as (\ref{cc}). 


\end{document}